\documentclass[12pt]{article}
\usepackage[dvips]{graphicx}
\makeatletter \@addtoreset{equation}{section}

\def\lsim{\mathrel{\rlap{\lower4pt\hbox{\hskip1pt$\sim$}}
    \raise1pt\hbox{$<$}}}      
\def\gsim{\mathrel{\rlap{\lower4pt\hbox{\hskip1pt$\sim$}}
    \raise1pt\hbox{$>$}}}      
\def\g2{ GeV$^2$}
\def\ie{\hbox{\it i.e. }}
\def\eg{\hbox{\it e.g. }}
\def\etal{\hbox{\it et al. }}

\begin{document}

\bigskip
\rightline{LYCEN2001-35 }

\bigskip
\bigskip

\begin{center}
{\Large \bf Regge models of the proton structure function\\ with and
without hard Pomeron:\\ a comparative analysis}
\bigskip
\bigskip
\end{center}

\begin{center}

{\large {
P. Desgrolard $^{a,}$\footnote{E-mail:
desgrolard@ipnl.in2p3.fr},
E. Martynov $^{b,}$\footnote{E-mail:
martynov@bitp.kiev.ua} }}

\end{center}

\bigskip
\bigskip

\noindent
$^a$ Institut de Physique Nucl\'eaire de Lyon, IN2P3-CNRS et
Universit\'e Claude Bernard, 43 boulevard du 11 novembre 1918, F-69622
Villeurbanne Cedex, France

\noindent $^b$ Bogolyubov Institute for Theoretical Physics, National
Academy of Sciences of Ukraine, 03143 Kiev-143, Metrologicheskaja 14b,
Ukraine

\bigskip

\bigskip

\begin{center}
\begin{minipage}[t]{13.0cm}
\noindent {\bf Abstract} A comparative phenomenological analysis of Regge
models with and without a hard Pomeron component is performed using a
common set of recently updated data. It is shown that the data at small
$x$ do not indicate explicitly the presence of the hard Pomeron. Moreover,
the models with two soft-Pomeron components (simple and double poles in
the angular momentum plane) with trajectories having intercept equal one
lead to the best description of the data not only at $W>3$ GeV and at
small $x$ but also at all $x\leq 0.75$ and $Q^{2}\leq 30000$ GeV$^{2}$.
\end{minipage}
\end{center}

\bigskip
\section{\Large{Introduction}}
It can be asserted confidently that Regge theory \cite{Collins} is one of
the most successful approaches to describe high energy scattering of
hadrons. Since some of important ingredients of amplitudes such as vertex
functions or couplings cannot be calculated (derived) theoretically, a
number of models are based on additional assumptions. Concerning the
leading Regge singularity, the Pomeron, even its intercept is a subject of
lively discussions. Moreover, the proper Regge models as well as the
models inspired by QCD or by other approaches, having elements of Regge
approach, are more or less successful when applied to processes induced by
photons (for an obviously incomplete list, see
\cite{GLR-83}-\cite{D-Lnew}).

Two methods are currently used to construct a phenomenological Pomeron
model for pure hadronic amplitudes. In the first one, the Pomeron is
supposed to be a simple pole in the angular momentum ($j$-) plane, with
intercept $\alpha_{P}(0)>1$. This property is necessary to explain the
observed growth of the total cross-sections with energy. Then, such a
Pomeron must be unitarized because it violates unitarity. In the second
approach, the amplitude is constructed, from the start, in accordance
with general requirements imposed by unitarity and analyticity. Here
Pomeron has $\alpha_{P}(0)=1$ and must be a singularity harder than the
simple pole is (again because of the rising cross-sections).

The hypothesis of Pomeron with $\alpha_{P}(0)>1$ (called sometimes
"supercritical" Pomeron) has a long history (see for example \cite{ChWu});
it is supported presently by perturbative QCD where BFKL Pomeron
\cite{BFKL} has $\Delta_{P}=\alpha_{P}(0)-1\approx 0.4$ in the leading
logarithmic approximation (LLA). However, the next correction to
$\Delta_{P}$ LLA is large and negative~\cite{FadLip}, the further
corrections being unknown yet. As a consequence, the intercept of the
Pomeron is usually determined phenomenologically from the experimental
data. In their popular supercritical Pomeron model, Donnachie and
Landshoff \cite{D-Lhadr} found $\alpha_{P}(0)=1.08$ from the data on
hadron-hadron and photon-hadron total cross-sections. When the model was
applied in deep inelastic scattering, namely to the proton structure
functions, the authors needed to add a second Pomeron, "hard", (in
contrast with the first one called "soft" Pomeron, because of its
intercept near 1), with a larger intercept $\alpha_{h P}(0)\approx 1.4$
\cite{D-L2pom,D-Lnew}.

At the same time, the detailed comparison~\cite{DGLM}-\cite{DGMP-ed}
of various models of Pomeron
with the data on total cross-section shows that
a better description (less value of $\chi^{2}$ and more stable values of
fitted parameters when the minimal energy of the data set is varying) is
achieved in alternative models with Pomeron having intercept one, but
being a harder $j$-singularity, for example, a double pole. Thus, the Soft
Dipole Pomeron (SDP) model was generalized for the virtual photon-proton
amplitude and applied to the proton structure function (SF) in a wide
kinematical region of deep inelastic
scattering \cite{DLM-dpom}. This model also has two Pomeron components,
each of them with intercept $\alpha_{P}(0)=1$, one is a double pole and
the other one is a simple pole.

Recent measurements of the SF have become available, from H1 \cite{ad00}
and ZEUS \cite{br00} collaborations ; they complete or correct the
previous data near the HERA collider~\cite{ah95}-\cite{ad99} and
\cite{de96}-\cite{br99} and from fixed target experiments~
\cite{ar97}-\cite{pdg}. They have motivated us to test and compare
above mentioned Pomeron models of the proton structure function
$F_2(x,Q^2)$ for the widest region of $Q^2$ and $x$.

In this paper, we would like to determine how is crucial or no the
existence of a hard Pomeron component (having in mind the previous
successes of the Soft Dipole Pomeron model without a hard Pomeron
component). We support the point of view that Pomeron is an universal
Reggeon: only the vertex functions are different with different processes.
That means that the Pomeron trajectory (or trajectories in the case of two
components) could not depend on the external particles \ie on the
virtuality $Q^{2}$ of photon in DIS. This circumstance dictates partially
the choice of the models under consideration. Our aim is to propose
a detailed quantitative comparison of some models, satisfying the
hypothesis of universality, with and without a hard Pomeron.

Details on the fitting procedure, particularly on the choice of
experimental data, are given in the next section. In Sect.3, the proposed
models are defined (or redefined), their comparison is performed in two
steps~: the low $x$ analysis allows to select the best ones kept in the
extended $x$-range.

\bigskip
\section{Fitting procedure~: details}
The choice of a data set may have crucial consequences in definitive
conclusions of any analysis. Thus, a set including the most recent and
older data has been used in the fits of the models of the proton SF. These
updated data are listed and referenced in Table~\ref{Expdata}. We have
fitted the models in three kinematic regions: A, B and C.
\begin{equation}\label{setA}
W>6\ \mbox{GeV}, \quad x\leq 0.07, \quad Q^{2}_{max}=3000\ \mbox{GeV}^{2},
\qquad \mbox{Region A}.
\end{equation}
\begin{equation}\label{setB}
W>3\ \mbox{GeV}, \quad x\leq 0.07, \quad Q^{2}_{max}=3000\ \mbox{GeV}^{2},
\qquad \mbox{Region B}
\end{equation}
\begin{equation}\label{setC}
W>3\ \mbox{GeV}, \quad x\leq 0.75, \quad Q^{2}_{max}=3000\ \mbox{GeV}^{2},
\qquad \mbox{Region C}
\end{equation}
The determination of the regions A (with 797 points) and B (with 878 points)
is arbitrary enough, especially concerning the upper limit for $x$,
aiming to select "small" $x$.

The second region (B) is the extension of region A for $W>3$ GeV. One can
see from Table~\ref{Expdata} that the difference between both comes mainly
from the added data on the cross-section $\sigma_{tot}^{\gamma p}$, when
we are going from A to B.

We remark that the pure hadronic cross-sections data at $\sqrt{s}\geq 5$
GeV are described well by the Dipole Pomeron~\cite{CEKLT,DGMP-ed},
whereas the physical threshold for $NN$ interaction is $\sqrt{s_{NN}}\sim 2$
GeV. For $\gamma N$ interaction the threshold is lower,
$\sqrt{s_{\gamma N}}\equiv {W_{\gamma N}}\sim 1$ GeV. Thus one can expect
a good description of the low $W$-data at least within the Soft Dipole
Pomeron model.
\begin{table}[ht]
  \centering
  \caption{Observables sets used in the fitting procedure
(note that the mentioned year does not correspond to the data-taking
period, but rather to the final publication. For description of the
different regions, see the text.}
\label{Expdata}
{\small
\begin{displaymath}
\begin{array}{|cc|c|c|c|}
\hline {\rm Observable} & & {\rm Region\ A } ({\rm A}_{1})& {\rm Region\ B
} ({\rm B}_{1})
& {\rm\ Region\ C} \\
{\rm Exp.-year\ of\ pub., }&{\rm \ Ref} & {\rm Nb\,points}&
{\rm Nb\,points}&{\rm Nb\,points} \\
\hline \hline
 F_2^p                      & & & &   \\
{\rm H1-1995}      \ \ & \cite{ah95}&  85  &  85  &  93 \\
{\rm H1-1996}      \ \ & \cite{ai96}&  37  &  37  &  41 \\
{\rm H1-1997}      \ \ & \cite{ad97}&  21  &  21  &  21 \\
{\rm H1-2000}      \ \ & \cite{ad99}&  51  &  51  & 111 \\
{\rm H1-2001}      \ \ & \cite{ad00}& 127  & 127  & 133 \\
{\rm ZEUS-1996}    \ \ & \cite{de96}& 153  & 153  & 186 \\
{\rm ZEUS-1997}    \ \ & \cite{br97}&  34  &  34  &  34 \\
{\rm ZEUS-1999}    \ \ & \cite{br99}&  44  &  44  &  44 \\
{\rm ZEUS-2000}    \ \ & \cite{br00}&  70  &  70  &  70  \\
{\rm NMC-1997}     \ \ & \cite{ar97}&  59  &  65  & 156 \\
{\rm E665-1996}    \ \ & \cite{ad96}&  80  &  80  &  91 \\
{\rm SLAC-1990/92} \ \ & \cite{slac}&   0  & 7(0) & 136 \\
{\rm BCDMS-1989}   \ \ & \cite{be89}& 5(0) & 5(0) & 175 \\
\hline
\sigma_{tot}^{\gamma ,p} \ \ &  &  & &\\
{\rm 1975/78;ZEUS-1994;H1-1995} & \cite{pdg} & 31 & 99 & 99  \\
\hline
{\rm Total}              &   & 797 (792)& 878 (866)& 1390  \\
\hline
\end{array}
\end{displaymath}
}
\end{table}

Running a few steps forward we should note that there are a few data
points from the fixed target experiments \cite{slac}, \cite{be89} in the
above mentioned regions A (5 points of BCDMS experiment) and B (5 points
of BCDMS and 7 points of SLAC experiments) that lead to some problems in
the fit. Firstly, they contribute to the $\chi^{2}$ noticeably more than
the other points do. Secondly, an analysis of all the models we consider
here shows that they destroy the stability of the parameters values when
one goes from region A to region B. The problems disappear if these 12
points are eliminated from our fit. Possibly, at small $x$, there is a
small inconsistency (due to normalization ?) between the experiments. In
the following, we present the detailed results of a fit without these
points (the corresponding data sets are noted as A$_{1}$ and B$_{1}$),
but we give also the values of $\chi^{2}$ for the full data sets, A and B.

The third region (C) includes all data listed in Table~\ref{Expdata}. The
relative normalization among all the experimental data sets has been fixed
to 1. Following the suggestion from~\cite{ad99}, some data from
~\cite{ai96} are considered as obsolete and superseded. They correspond to
($Q^2\ge 250$ \g2 , for all $x$), ($Q^2= 200$ \g2 , for $x<0.1$) and
($Q^2= 150$ \g2 , for $x<0.01$). We also cancelled the ancient values
(with moderate $Q^2\le 150 $ \g2~: 88 from~\cite{ai96} and 23
from~\cite{ad97}) which have been duplicated in the more recent high
precision measurements~\cite{ad00}. We have excluded the whole domain
$Q^{2}\geq 5000$ GeV$^{2}$ from the fit (19 data points from~\cite{ad99}
and 2 from~\cite{de96}), because the difference (experimentally observed)
between $e^{-}p$ and $e^{+}p$ results cannot be (and should not be)
explained by Pomeron + $f$ exchange. No other filtering of the data has
been performed. Experimental statistical and systematic errors are added
in quadrature.

As usual, we "measure" the quality of agreement of each model
with experimental data by the
$\chi^2$, minimized using the MINUIT computer code.
The ensuing determination of the free parameters is associated with the
corresponding one-standard deviation errors.
The results are displayed below~\footnote{
In following Tables 2-7 the values
of parameters and errors are presented in the form given by MINUIT, not
rounded.}.

\bigskip
\section{Regge models in Deep Inelastic Scattering and \\
phenomenological analysis} We stress again that there are numerous models
for the proton SF, inspired by a Regge approach, which describe more or
less successfully the available data on the SF in a wide region of $Q^2$
and $x$. Here, we consider two of them (and their modifications): the
two-Pomeron model of Donnachie and Landshoff \cite{D-L2pom} and the Soft
Dipole Pomeron model \cite{DLM-dpom}, incorporating explicitly the ideas
of universality for a Reggeon contribution (in the Born approximation) and
of $Q^{2}$-independent intercepts for Pomeron and $f$-Reggeon
trajectories. We compare these models using the above common set of
experimental data.

\medskip
\subsection{Kinematics}
We use the standard kinematic variables to describe deep inelastic
scattering (DIS)~:
\begin{equation}\label{ep eX}
 e(k) \ +\ p(P) \ \to\ e(k')\ +\ X\ ,
\end{equation}
 where $ k, k', P$ are the four-momenta of the incident electron,
scattered electron and incident proton. $ Q^2$ is the negative squared
four-momentum transfer carried by the virtual exchanged photon
(virtuality)
\begin{equation}\label{defQsq}
Q^2 \ =\ -q^2\ =\ -(k-k')^2\ .
\end{equation}
$x$ is the Bj\"orken variable
\begin{equation}\label{defx}
x\ =\ {Q^2\over 2P\cdot q} \ ,
\end{equation} $W$ is the center of mass
energy of the ($\gamma^*, p$) system, related to the above variables by
\begin{equation}\label{defW}
 W^2 \ =\ (q+P)^{2} \ =\ Q^2{1-x\over x}+m_p^2 \ ,
 \end{equation}
 with $m_p$
being the proton mass.

\medskip
\subsection{Soft and Hard Pomeron models at small $x$.}

\smallskip
\subsubsection{\bf Soft + Hard Pomeron (S+HP) model.}
Considering the two-Pomeron model of Donnachie and Landshoff (D-L), we use
a recently published variant~\cite{D-L2pom}~\footnote{When the present
paper was practically finished, an other variant~\cite{D-Lnew} appeared
with a slightly changed soft Pomeron term and additional factors
$(1-x)^{b}$ in each term . We repeated our calculations for this new
version however we failed to obtain $\chi^{2}/d.o.f.<1.5$ even for region
A$_{1}$ if the soft pomeron term (\ref{DLorFs}) does not have square root
factor.} and write the proton SF as the sum of three Regge~contributions:
a hard and a soft Pomeron and an $f$-Reggeon
\begin{equation}\label{DLorF2}
F_{2}(x,Q^{2})=F_{hard}+F_{soft}+F_{f}\ ,
\end{equation}
where
\begin{equation}\label{DLorFh}
F_{hard}=C_{h}\left( \frac{Q^{2}}{Q^{2}+Q^{2}_{h}}\right)^{1+\epsilon_{h}}
\left(1+\frac{Q^{2}}{Q^{2}_{h}}\right)^{\frac{1}{2}\epsilon_{h}}
\left(\frac{1}{x}\right)^{\epsilon_{h}},
\end{equation}
\begin{equation}\label{DLorFs}
F_{soft}=C_{s}\left( \frac{Q^{2}}{Q^{2}+Q^{2}_{s}}\right)^{1+\epsilon_{s}}
\left(1+\sqrt{\frac{Q^{2}}{Q^{2}_{s\,0}}}\right)^{-1}
\left(\frac{1}{x}\right)^{\epsilon_{s}},
\end{equation}
\begin{equation}\label{DLorFf}
F_{f}=C_{f}\left( \frac{Q^{2}}{Q^{2}+Q^{2}_{f}}\right)^{\alpha_{f}(0)}
\left(\frac{1}{x}\right)^{\alpha_{f}(0)-1}
\end{equation}
with the cross-section (we approximate the total cross-section as the
transverse one)
\begin{equation}\label{DLorSig}
\sigma_{T}^{\gamma p}(W^2)=\frac{4\pi^{2}\alpha}{Q^{2}}F_{2}(x,Q^{2})
\bigg |_{Q^{2}=0}=4\pi^{2}\alpha\sum\limits_{i=h,s,f}
\frac{C_{i}}{(Q_{i}^{2})^{1+\epsilon_{i}}}(W^{2}-m_{p}^{2})^{\epsilon_{i}}.
\end{equation}
where $\epsilon_{f}=\alpha_{f}(0)-1.$

We show in Table 2 results of the fit  performed in the regions
A$_{1}$ and B$_{1}$.

\begin{table}[ht]
\caption{Parameters of the "Soft + Hard Pomerons" model~\cite{D-L2pom}
obtained from our fits in the regions A$_{1}$ and B$_{1}$ }\label{DLorpar}
\begin{center}
\begin{tabular}{|l||c|c||c|c|}
\hline Parameter & \multicolumn{2}{c||}{ Fit A$_{1}$ ($W\geq 6$ GeV)} &
\multicolumn{2}{c|}{ Fit B$_{1}$ ($W\geq 3$ GeV)}\\
\cline{2-5}
&value & $\pm$ error & value & $\pm$ error \\
\hline\hline
 $C_{h}$                       &  .460299E-01 & .139016E-02 & .479219E-01 & .334416E-02\\
 $\epsilon_{h}$                &  .435721E+00 & .417590E-02 & .431656E+00 & .945276E-02\\
 $Q^{2}_{h}$ (GeV$^{2}$)       &  .101656E+02 & .253932E+00 & .985373E+01 & .479673E+00\\
\hline\hline
 $C_{s}$                       &  .356527E+00 & .353337E-02 & .354170E+00 & .777590E-02\\
 $\epsilon_{s}$                &  .891515E-01 & .138958E-02 & .900833E-01 & .319777E-02\\
 $Q^{2}_{s}$ (GeV$^{2}$)       &  .667118E+00 & .779835E-02 & .672313E+00 & .213077E-01\\
 $Q^{2}_{s\,0}$ (GeV$^{2}$)    &  .129234E+03 & .143884E+02 & .912972E+02 & .227659E+02\\
\hline\hline
 $C_{f}$                       &  .340843E-02 & .282839E-03 & .475168E-01 & .118216E-01\\
 $\alpha_{f}(0)$               &  .572827E+00 & .578989E-02 & .602908E+00 & .268310E-01\\
 $Q^{2}_{f}$ (GeV$^{2}$)       &  .351976E-04 & .487792E-05 & .540418E-02 & .228127E-02\\
\hline\hline
$\chi^{2}/d.o.f.$ & \multicolumn{2}{c||} {1.089} &
 \multicolumn{2}{c|} {1.176}\\
 \hline
\end{tabular}
\end{center}
\end{table}
In order to take full advantage of the parametrization, but in contradiction
with the original more economic suggestion of D-L, we allowed for the
intercepts of the Soft Pomeron and $f$- Reggeon to be free.

In both regions, the values of $Q_{f}^{2}$ are found too small. If we put
the low limit for this parameter at 0.076 GeV$^{2}$($\approx
4m_{\pi}^{2}$, minimal physical threshold in $t$-channel), then
$\chi^{2}/d.o.f.$ increases a little up to 1.192 in the fit A$_{1}$ and up
to 1.283 in the fit B$_1$.

If the above mentioned 12 BCDMS and SLAC points are taken into account
then we obtain
$$
\chi^{2}/d.o.f.=1.097 \qquad \mbox{in region A},
$$
$$
\chi^{2}/d.o.f.=1.504 \qquad \mbox{in region B}
$$
with free intercepts of Pomeron and $f$-Reggeon.

One can see that decreasing the minimal energy of the data set always
leads to a deterioration of the fit.

\smallskip
\subsubsection{\bf Soft Dipole Pomeron (SDP) model}\label{SectSDP}
Defining the Dipole Pomeron model for DIS, we start from the expression
connecting the transverse cross-section of $\gamma^*p$ interaction to the
proton structure function $F_2$ and the optical theorem for forward
scattering amplitude~\footnote{ Note the $8\pi$ factor in
the optical theorem non included in~\cite{DLM-dpom}. }
\begin{equation}
 \sigma_{T}^{\gamma^*p}=8\pi\Im
m A(W^2,Q^2;t=0)=\frac{4\pi^2\alpha}{Q^2(1-x)}(1+4m_p^2x^2/Q^2)
F_2(x,Q^2) ;
\end{equation}
the longitudinal contribution to the total cross-section,
$\sigma_L^{\gamma^*p}=0$ is assumed. Though we consider in this Subsection
only small $x$ we give here the complete parameterization \cite{DLM-dpom}
valid also at large values of $x$; it will be fully exploited in the next
Section. The forward scattering at $W$ far from the $s$-channel threshold
$W_{th}=m_p$ is dominated by the Pomeron and the $f$-Reggeon
\begin{equation}\label{ampl}
 A(W^2,t=0;Q^2)=P(W^2,Q^2)+f(W^2,Q^2),
\end{equation}
\begin{equation}\label{f-term}
 f(W^2,Q^2)=iG_f(Q^2)(-iW^2/m_p^2)^{\alpha_f(0)-1}(1-x)^{B_f}.
\end{equation}
\begin{eqnarray}
G_f(Q^2)=\frac{C_f}{\left(1+Q^2/Q_{f}^2
\right)^{D_f(Q^2)}},\label{G-f} \\
D_f(Q^2)=d_{f\infty}+\frac{d_{f0}-d_{f\infty}}{1+Q^2/
Q_{fd}^2},\label{D-qf}
\end{eqnarray}
\begin{equation}\label{B-xf}
B_f(Q^2)=b_{f\infty}+\frac{b_{f0}-b_{f\infty}}{1+Q^2/Q^2_{fb}}.
\end{equation}
As for the Pomeron contribution, we take it in the two-component form
\begin{equation}\label{pom-term}
P(W^2,Q^2)=P_1+P_2,
\end{equation}
with
\begin{eqnarray}
P_1&=&iG_1(Q^2)\ell n(-iW^{2}/m_{p}^{2})(1-x)^{B_1(Q^2)},\label{p-1} \\
P_2&=&iG_2(Q^2)                         (1-x)^{B_2(Q^2)},\label{p-2}
\end{eqnarray}
where
\begin{eqnarray}
G_i(Q^2)&=&\frac{C_i}{\left(1+Q^2/Q_{i}^2
\right)^{D_i(Q^2)}}, \qquad i=1,2,\label{G-ip} \\
\qquad D_i(Q^2)&=&d_{i\infty}+\frac{d_{i0}-d_{i\infty}}{1+Q^2/ Q_{id}^2},
\qquad i=1,2,\label{D-qp}
\end{eqnarray}
\begin{equation}\label{B-xp}
B_i(Q^2)=b_{i\infty}+\frac{b_{i0}-b_{i\infty}}{1+Q^2/Q^2_{ib}}, \qquad
i=1,2.
\end{equation}
We would like to comment the above expressions, especially the powers
$D_{i}$ and $B_{i}$ varying smoothly between constants when $Q^{2}$ goes
from $0$ to $\infty$. In spite of an apparently cumbersome form they are a
direct generalization of the exponents $d$ and $b$ appearing in each term
of the simplest parametrization of the $\gamma^*p$-amplitude
$$
G(Q^2)=\frac{C}{(1+Q^2/Q_0^2)^d}\qquad \mbox{and}\qquad (1-x)^b\ .
$$
Indeed, a fit
to experimental data shows unambiguously that the parameters $d$ and $b$
should depend on $Q^2$.

At small $x\leq 0.07$, which are under interest now, it is not necessary
to keep factors $(1-x)^{B_{i}}$, significant only when $x$ gets near 1, in
(\ref{f-term},\ref{p-1},\ref{p-2}), with $B_{i}=B_{i}(Q^2)$. In order to
exclude in the expression for $F_{2}$ (rather than for $
\sigma_{T}^{\gamma^*p}$) any factors $(1-x)$, we should fix $B_{i}=-1$ in
the above equations. In this case the S+HP and the SDP models can be
compared for small $x$ under similar conditions.

The results of fitting the data in the regions A$_1$ and $B_1$ are given in
Table~\ref{DPpar}.

{\renewcommand{\arraystretch}{1.2}%
\begin{table}[ht]
\caption{Parameters fitted in the Soft Dipole Pomeron
model~\cite{DLM-dpom} simplified in the small-$x$ regions A$_{1}$ and
B$_{1}$ .} \label{DPpar}
{\footnotesize
\begin{center}
\begin{tabular}{|l||c|c||c|c|}
\hline Parameter & \multicolumn{2}{c||}{ Fit A$_{1}$ ($W\geq 6$ GeV)} &
\multicolumn{2}{c|}{ Fit B$_{1}$ ($W\geq 3$ GeV)}\\
\cline{2-5}
&value & $\pm$ error & value & $\pm$ error \\
\hline \hline
$C_1$ (GeV$^{-2}$)               &  .222698E-02&   .494472E-05&   .222972E-02&   .474366E-05  \\
$Q_1^2$ (GeV$^2)$                &  .860804E+01&   .213165E-01&   .823921E+01&   .207764E-01  \\
$Q_{1d}^2$ (GeV$^2)$             &  .126690E+01&   .694433E-02&   .123648E+01&   .711041E-02  \\
$d_{1\infty}$                    &  .124426E+01&   .301742E-02&   .124032E+01&   .298242E-02  \\
$d_{10}$                         &  .986652E+01&   .391815E-01&   .941385E+01&   .384359E-01  \\
\hline \hline
$C_2$ (GeV$^{-2}$)               &$-$.893679E-02&  .228570E-04&$-$.889722E-02&   .228190E-04  \\
$Q_2^2$ (GeV$^2)$                &  .198085E+02&   .496695E-01&   .189066E+02&   .486844E-01  \\
$Q_{2d}^2$ (GeV$^2)$             &  .947165E+00&   .687587E-02&   .105738E+01&   .820888E-02  \\
$d_{2\infty}-d_{1\infty}$ (fixed)&  .000000E+00&   .000000E+00&   .000000E+00&   .000000E+00  \\
$d_{20}$                         &  .132941E+02&   .939077E-01&   .109375E+02&   .818441E-01  \\
\hline \hline
$\alpha_f(0)$ \ (fixed)          &  .785000E+00&   .000000E+00&   .785000E+00&   .000000E+00  \\
$C_f$ (GeV$^{-2}$)               &  .294850E-01&   .893680E-04&   .293259E-01&   .774985E-04  \\
$Q_f^2$ (GeV$^2)$                &  .182986E+02&   .985102E-01&   .176838E+02&   .958663E-01  \\
$Q_{fd}^2$ (GeV$^2)$             &  .616179E+00&   .495589E-02&   .659804E+00&   .541872E-02  \\
$d_{f\infty}$                    &  .134520E+01&   .401477E-02&   .135104E+01&   .410773E-02  \\
$d_{f0}$                         &  .404273E+02&   .267141E+00&   .357568E+02&   .236524E+00  \\
\hline\hline $\chi^{2}/d.o.f.$ & \multicolumn{2}{c||} {0.911} &
 \multicolumn{2}{c|} {0.948}\\
\hline
\end{tabular}
\end{center}
}
\end{table}
}
The intercept of $f$-Reggeon is then fixed at the value
$\alpha_{f}(0)=0.785$ obtained \cite{DGMP-ed} from the fit to hadronic
total cross-sections.

One can see from this table that the quality of the data description in
the Soft Dipole Pomeron model is quite high. Furthermore, the values of
the fitted parameters are close in both regions. Thus we claim a good
stability of the model when the minimal energy $W$ of the data set is
varying.

Moreover, and to enforce this statement, we have investigated the ability
of the SDP model to describe data in other kinematical regions namely with
"small" $x\leq 0.1$ and $Q^{2}_{max}=3000$ Gev$^2$. Results follow
$$
\chi^{2}/d.o.f.=0.978 \qquad \mbox{if} \qquad W\geq 6 \mbox{ GeV},
$$
$$
\chi^{2}/d.o.f.=1.014 \qquad \mbox{if} \qquad W\geq 3 \mbox{ GeV}.
$$
Parameters are stable again and are not strongly modified from those in
Table~\ref{DPpar} for the regions A$_{1}$ and B$_{1}$.

If BCDMS and SLAC points are included in the fits the following
results are obtained for $x\leq 0.07$
$$
\mbox{Region A:} \qquad \chi^{2}/d.of.=0.936 ,
$$
$$
\mbox{Region B:} \qquad \chi^{2}/d.of.=1.021.
$$
However, as already noted, some of the fitted parameters are not stable
under transition from region A to region B (in the present case, mainly
the parameters $d_{i0}$ are concerned).

\smallskip
\subsubsection{\bf Modified two-Pomeron (Mod2P) model}\label{SectMod2P}
We already noted elsewhere~\cite{DGLM,DGMP-ed,MAQM} a very interesting
phenomenological fact which occurs for total cross-sections. If a constant
term (or a contribution from a Regge pole with intercept one) is added to
the ordinary "supercritical" Pomeron with $\alpha_{P}(0)=1+\epsilon$ (for
example in the popular Donnachie-Landshoff model~\cite{D-Lhadr})
the fit to the available data leads to a very small value of $\epsilon
\sim 0.001$ and to a negative sign of the new constant term. This is valid
when $pp$ and $\bar pp$ total cross-sections are considered as well as
when all cross-sections, including $\sigma_{tot}^{\gamma p}$ and
$\sigma_{tot}^{\gamma \gamma}$, are taken into account. Due to this small
value of $\epsilon$ one can expand the factor $(-is/s_{0})^{\epsilon}$,
entering in the supercritical Pomeron, keeping only two first terms and
obtain, in fact, the Dipole Pomeron model. We would like to emphasize that
the resulting parameters in such a modified Donnachie-Landshoff model for
total cross-sections are very close to those obtained in the Dipole
Pomeron model.

It has been demonstrated above that SDP model for $F_{2}(x,Q^{2})$,
simplified for low $x$, describes well (even better than S+HP model does)
DIS data in a wide region of $Q^{2}$. A natural question arises~: does
such a situation remain possible for $\sigma_{T}^{\gamma^{*}p}$ or for the
proton structure function at any $Q^{2}$~? In what follows, we suggest a
modification of the model defined by (\ref{DLorF2}-\ref{DLorFf}) and
argue that answer on the above question is positive.

In fact, we consider the original S+HP model with $\epsilon_{h}=0$
modifying only residues and redefining the coupling constants~\footnote{We
change also index "h" for "0" because now the term $F_{0}$ is no more a
contribution of the hard Pomeron with high intercept. } to have for
the cross-section the expression
\begin{equation}
\label{sigDLmod} \sigma ^{\gamma p}_{T}(W^2)=4\pi^{2}\alpha \left \{
\frac{C_{0}}{\epsilon}+
\frac{C_{s}}{\epsilon}\left(\frac{W^{2}}{m_{p}^{2}}-1\right)^{\epsilon}+
C_{f}\left(\frac{W^{2}}{m_{p}^{2}}-1\right)^{\alpha_{f}(0)-1}\right \}.
\end{equation}
$\epsilon$ is inserted in the denominators in order to avoid
large values of $C_{0}$ and $C_{s}$ when $\epsilon\ll 1$ (this case
occurs in the fit).

Thus we write
\begin{equation}\label{DLmod}
F_{2}(x,Q^{2})=F_{0}+F_{s}+F_{f}
\end{equation}
where
\begin{equation}\label{DLmod0}
F_{0}=\frac{C_{0}Q^{2}_{0}}{\epsilon }\left(
\frac{Q^{2}}{Q^{2}+Q^{2}_{0}}\right)
\left(1+\frac{Q^{2}}{Q^{2}_{0\,1}}\right)^{d_{0}},
\end{equation}
\begin{equation}\label{DLmods}
F_{s}=\frac{C_{s}Q^{2}_{s}}{\epsilon (
m_{p}^{2}/Q^{2}_{s})^{\epsilon}}\left(
\frac{Q^{2}}{Q^{2}+Q^{2}_{s}}\right)^{1+\epsilon}
\left(1+\frac{Q^{2}}{Q^{2}_{s\,1}}\right)^{d_{s}}
\left(\frac{1}{x}\right)^{\epsilon},
\end{equation}
\begin{equation}\label{DLmodf}
F_{f}=\frac{C_{f}Q^{2}_{f}}{(m_{p}^{2}/Q^{2}_{f})^{\alpha_{f}(0)-1}}\left(
\frac{Q^{2}}{Q^{2}+Q^{2}_{f}}\right)^{\alpha_{f}(0)}
\left(1+\frac{Q^{2}}{Q^{2}_{f\,1}}\right)^{d_{f}}
\left(\frac{1}{x}\right)^{\alpha_{f}(0)-1}.
\end{equation}

 The values of the free parameters and $\chi^{2}$ are given in
Table~\ref{DLmodpar}.

{\renewcommand{\arraystretch}{1.2}%
\begin{table}[ht]
\caption{Values of the fitted parameters in the Modified two-Pomeron
model}\label{DLmodpar} {\footnotesize
\begin{center}
\begin{tabular}{|l||c|c||c|c|}
\hline Parameter & \multicolumn{2}{c||}{ Fit A$_{1}$ ($W\geq 6$ GeV)} &
\multicolumn{2}{c|}{ Fit B$_{1}$ ($W\geq 3$ GeV)}\\
\cline{2-5}
&value & $\pm$ error & value & $\pm$ error \\
\hline\hline
 $C_{0}$ (GeV$^{-2}$)      &$-$.192614E+00&.347649E-05  &$-$.192597E+00&.341593E-05\\
 $Q^{2}_{0}$ (GeV$^{2}$)   &   .103160E+01&.240636E-04  &   .103160E+01&.240085E-04\\
$Q^{2}_{0\ 1}$ (GeV$^{2}$) &   .120880E+01&.991056E-04  &   .120742E+01&.986103E-04\\
 $d_{0}$                   &   .287793E+00&.479038E-05  &   .288178E+00&.477790E-05\\
\hline\hline
 $C_{s}$ (GeV$^{-2}$)      &   .191862E+00&.345642E-05  &   .191835E+00&.339575E-05\\
 $\epsilon $ (fixed)       &   .101300E-02&.000000E+00  &   .101300E-02&.000000E+00\\
 $Q^{2}_{s}$ (GeV$^{2}$)   &   .980419E+00&.227654E-04  &   .980571E+00&.227169E-04\\
 $Q^{2}_{s\ 1}$ (GeV$^{2}$)&   .101808E+01&.825788E-04  &   .101742E+01&.822231E-04\\
 $d_{s}$                   &   .287548E+00&.469776E-05  &   .287925E+00&.468653E-05\\
\hline\hline
 $C_{f}$ (GeV$^{-2}$)      &   .230047E+01&.828553E-02  &   .233104E+01&.753061E-02\\
 $\alpha_{f}(0)$ (fixed)   &   .789500E+00&.000000E+00  &   .789500E+00&.000000E+00\\
 $Q^{2}_{f}$ (GeV$^{2}$)   &   .102321E+01&.622960E-02  &   .987669E+00&.571238E-02\\
 $Q^{2}_{f\ 1}$ (GeV$^{2}$)&   .703071E+01&.128642E+00  &   .666517E+01&.119554E+00\\
 $d_{f}$                   &   .317443E+00&.130398E-02  &   .319035E+00&.128344E-02\\
\hline\hline $\chi^{2}/d.o.f.$ & \multicolumn{2}{c||} {0.959} &
 \multicolumn{2}{c|} {0.996}\\
\hline
\end{tabular}
\end{center}
}
\end{table}
}

One can see in Table~\ref{DLmodpar} that $d_{0}>d_{s}$ and that $C_0$ is
negative. Consequently, at some high values of $Q^{2}>Q_{m}^{2}(x)$, the
SF (\ref{DLmod}) turns out to become negative. Numerically the minimal
value of $Q^{2}_{m}$ where it occurs is \eg $Q_{m}^{2}\sim 4\cdot 10^{4}$
GeV$^{2}$ at $x\sim 0.05$. It is far beyond the kinematical limit $y
={Q^2\over x(s-m_p^2)}\le 1$, with $s-m_p^2\approx 4E_e E_p$, in terms of
the positron $E_e$ and proton $E_p$ beam energies of an ($ep$) collider.
For example, HERA measurements are presently restricted by $Q^2$(\g2 )
$\lsim 10^5 x$. Besides this, at a so high virtuality, one-photon
exchanges must be supplemented with other exchanges. On the other hand,
from a phenomenological point of view, a fit respecting the condition
$\delta=d_{s}-d_{0}\geq 0$ yields the lower limit $\delta =0$ and we
obtained then $\chi^{2}/d.o.f.\approx 1.057$ in the region A$_{1}$, better
than in the S+HP model with a hard Pomeron. Finally, the result could be
improved when replacing the constants $d_{i}$ by functions $D_{i}(Q^{2})$
such as (\ref{D-qf}), (\ref{D-qp}) in the SDP model. We do not consider
this possibility in order to avoid an extra number of parameters.

 For intercepts of Pomeron ($\epsilon$) and of $f$-Reggeon
($\alpha_{f}(0)$), the values obtained in~\cite{DGMP-ed}, in the case of
non degenerated and non universal SCP are taken and fixed, in accordance
with the idea of Reggeon universality (and because the data for
$\sigma_{tot}^{\gamma p}$ are insufficient to determine precisely and
simultaneously both the intercepts and the couplings).

For fits in the kinematical regions A and B (with BCDMS and SLAC points
included) we have
$$
\mbox{Region A:} \qquad \chi^{2}/d.of.=0.963,
$$
$$
\mbox{Region B:} \qquad \chi^{2}/d.of.=1.004.
$$

\smallskip
To complete the set of Regge models, we present now
an other modification of the Donnachie and Landshoff model. At the same
time, it can be considered as a generalization of the Soft Dipole Pomeron
model.
\subsubsection{\bf Generalized logarithmic Pomeron (GLP) model.}
\label{SectGLP} We have found in \cite{DLM-bx} a shortcoming of the SDP
model, relative to the logarithmic derivative $B_{x}=\partial \ell
nF_{2}(x,Q^{2})/\partial \ell n(1/x)$ at large $Q^{2}$ and small $x$.
Namely, in spite of a good $\chi^{2}$ in fitting the SF, theoretical
curves for $B_{x}$ are systematically lower than the data on this quantity
extracted from $F_{2}$. In our opinion, one reason might be a
insufficiently fast growth of $F_{2}$ with $x$ at large $Q^{2}$ and small
$x$ (the SDP model leads to a logarithmic behaviour in $1/x$) On the other
side, essentially a faster growth of $F_{2}$ (and consequently of $B_{x}$)
is, from a phenomenological point of view, a good feature of the D-L model.
However, this model violates the known Froissart-Martin bound on
the total cross-section of $\gamma^*p$ process which, as commonly believed,
should be valid at least for real photons.

Thus, we have tried to construct a model that incorporates a slow rise of
$\sigma_{T}^{\gamma p}(W^{2})$ and simultaneously a fast rise of
$F_{2}(x,Q^2)$ at large $Q^{2}$ and small $x$. We propose below a model
intended to link these desirable properties, being in a sense intermediate
between the Soft Dipole Pomeron model (\ref{ampl})--(\ref{B-xp}) and the
Modified two Pomeron (\ref{DLmod})--(\ref{DLmodf}) model. Again, as for
SDP, we give a parameterization valid for all $x$, without restriction.
\begin{equation}\label{LogG}
F_{2}(x,Q^{2})=F_{0}+F_{s}+F_{f},
\end{equation}
\begin{equation}\label{LogG-0}
F_{0}=C_{0}\frac{Q^{2}}{\left(1+Q^{2}/Q^{2}_{0}\right)^{d_{0}}}
(1-x)^{B_{0}(Q^{2})},
\end{equation}
\begin{equation}\label{LogG-s}
F_{s}=C_{s}\frac{Q^{2}}{(1+Q^{2}/Q^{2}_{s })^{d_{s}}}L(Q^{2},W^{2})
(1-x)^{B_{s}(Q^{2})},
\end{equation}
where
\begin{equation}\label{L(Q,W)}
L(Q^{2},W^2)=\ell n \left[1+\frac{a}{(1+Q^{2}/Q^{2}_{s\,0})^{d_{s0}}}
\left(\frac{Q^{2}}{xm_{p}^{2}}\right)^{\epsilon}\right]
\end{equation}
\begin{equation}\label{LogG-f}
F_{f}=C_{f}\frac{Q^{2}}{(1+Q^{2}/Q^{2}_{f})^{d_{f}}}
\left(\frac{Q^{2}}{xm_{p}^{2}}\right)^{\alpha_{f}(0)-1}(1-x)^{B_{f}(Q^{2})},
\end{equation}
where
\begin{equation}\label{LogG-B}
B_{i}(Q^{2})=b_{i\infty}+\frac{b_{i0}-b_{i\infty}}{1+Q^{2}/Q^{2}_{ib}},
\qquad i=0, s, f.
\end{equation}
A few comments on the above model are needed.
\begin{itemize}
\item
In the original D-L model the dependence on $x$ is in the form
$(Q^{2}/x)^{\epsilon}$ but with $(Q^{2})^{\epsilon}$ absorbed in a
coupling function $(Q^{2}/(Q^{2}+Q^{2}_{s}))^{1+\epsilon}$. The main
modification (apart from a replacement of a power dependence by a
logarithmic one) is that we inserted $(Q^{2})^{\epsilon}$ into "energy"
variable $Q^{2}/x$ and made it dimensionless. By a similar way we modified
the $f$-term.
\item
The new logarithmic factor in (\ref{LogG-s}) can be rewritten in the form
$$
L(Q^{2},W^2)=\ell n\left[1+\frac{a}{(1+Q^{2}/Q^{2}_{s0})^{d_{s0}}}\left(
\frac{W^{2}+Q^{2}}{m_p^{2}}-1\right)^{\epsilon}\right].
$$
At $Q^{2}=0$, we have $L(0,W^{2})=\ell n[1+a(W^{2}/m_p^{2}-1)^{\epsilon}]$
and consequently $L(0,W^{2})\approx \epsilon\ell n(W^{2}/m_p^{2})$ at
$W^{2}/m_p^{2}\gg 1$. Thus, $\sigma_{T}^{\gamma p}(W)\propto \ell n W^{2}$
at $W^{2}\gg m_p^{2}$. A similar behaviour can be seen at moderate $Q^{2}$
when the denominator is $\sim 1$. However at not very large
$W^{2}/m_p^{2}$ or at sufficient high $Q^{2}$ the argument of logarithm is
close to 1, and then
$$
L(Q^{2},W^2)\approx \frac{a}{(1+Q^{2}/Q^{2}_{s0})^{d_{s0}}}\left(
\frac{W^{2}}{m_p^{2}}-1\right)^{\epsilon}
$$
simulating a Pomeron contribution with intercept $\alpha_{P}(0)=1+\epsilon$.
\item
We are going to justify that, in spite of its appearance, the GLP model
cannot be treated as a model with a hard Pomeron, even when $\epsilon$
issued from the fit is not small. In fact, the power $\epsilon $ inside
the logarithm is NOT the intercept (more exactly is not
$\alpha_{P}(0)-1$). Intercept is defined as position of singularity of the
amplitude in $j$-plane at $t=0$. In our case, the true leading Regge
singularity is located exactly at $j=1$: it is a double pole due to the
logarithmic dependence. Let consider any fixed value of $Q^{2}$ and
estimate the partial amplitude with the Mellin transformation
$$
\begin{array}{ll}
\phi(j,t=0) & \sim
\int\limits_{W^{2}_{min}}^{\infty}dW^{2}\left (\frac{W^{2}}
{W^{2}_{min}}\right )^{-j}A(W^{2},0)\\
& \propto
\int\limits_{W^{2}_{min}}^{\infty}\frac{dW^{2}}{W^{2}}e^{-(j-1)\ell
n(W^{2}/W^{2}_{min})}\ell
n(1+a\frac{[(W^{2}+Q^{2})/m_{p}^{2}-1]^{\epsilon}}
{(1+Q^{2}/Q_{0}^{2})^{d_{s0}}}).
\end{array}
$$
One can see that the singularities of $\phi (j,0)$ are generated by a
divergence of the integral at the upper limit. To extract them we can put
the low limit large enough, say $W^{2}_{1}$. The remaining integral, from
$W^{2}_{min}$ to $W^{2}_{1}$, will only contribute to the non-singular
part of $\phi$. We can take $W^{2}_{1}$ so large to allow the
approximation $\ell n(1+a\frac{[(W^{2}+Q^{2})/m_{p}^{2}-1]^{\epsilon}}
{(1+Q^{2}/Q_{0}^{2})^{d_{s0}}})\approx \epsilon \ell n(W^{2}/m_{p}^{2})$.
In this approximation
$$
\phi(j,t=0)\propto \int\limits_{\zeta_{1}}^{\infty}d\zeta e^{-(j-1)\zeta
}\zeta \approx \frac{1}{(j-1)^{2}}\quad {\rm with }\quad \zeta_{1}=\ell
n(W^{2}_{1}/m_{p}^{2}).
$$
\item
Thus this model can be considered as a Dipole Pomeron model. In
order to distinguish between it and the Soft Dipole Pomeron model
presented in Section \ref{SectSDP}, we call this model as Generalized
Logarithmic Pomeron (GLP) model.

\end{itemize}

Performing fit in the regions A$_1$ and B$_1$, we fixed all $b_{i}=0$, as
required by the small $x$ approximation, $\alpha_P(0)$ as in SDP, and
obtained the results presented in Table~\ref{TLogGpar}.

{\renewcommand{\arraystretch}{1.2}%
\begin{table}[ht]
\caption{Values of the fitted parameters in the Generalized Logarithmic
Pomeron model, simplified for low $x$}\label{TLogGpar}
{\footnotesize
\begin{center}
\begin{tabular}{|l||c|c||c|c|}
\hline Parameter & \multicolumn{2}{c||}{ Fit A$_{1}$ ($W\geq 6$ GeV)} &
\multicolumn{2}{c|}{ Fit B$_{1}$ ($W\geq 3$ GeV)}\\
\cline{2-5}
&value & $\pm$ error & value & $\pm$ error \\
\hline\hline
 $C_{0}$ (GeV$^{-2}$)      &$-$.586790E+00 & .590162E-02 &$-$.622785E+00 &.573266E-02\\
 $Q^{2}_{0}$ (GeV$^{2}$)   &   .791206E+00 & .121230E-01 &   .790649E+00 &.113281E-01\\
 $d_{0}$                   &   .823491E+00 & .264806E-02 &   .823973E+00 &.251681E-02\\
\hline\hline
 $C_{s}$  (GeV$^{-2}$)     &   .591058E+00 & .341296E-02 &   .610089E+00 &.334329E-02\\
 $a $                      &   .792225E+00 & .912390E-02 &   .871875E+00 &.935499E-02\\
 $\epsilon $               &   .331868E+00 & .146863E-02 &   .318210E+00 &.138877E-02\\
 $Q^{2}_{s\,0}$ (GeV$^{2}$)&   .566155E+00 & .117036E-01 &   .512134E+00 &.101668E-01\\
 $d_{s\, 0}$               &   .654811E+00 & .276609E-02 &   .650015E+00 &.260637E-02\\
 $Q^{2}_{s}$ (GeV$^{2}$)   &   .447791E+00 & .620657E-02 &   .458533E+00 &.607371E-02\\
 $d_{s}$                   &   .541809E+00 & .227447E-02 &   .532853E+00 &.221010E-02\\
\hline\hline
 $C_{f}$  (GeV$^{-2}$)     &   .201820E+01 & .150381E-01 &   .198981E+01 &.120945E-01\\
 $\alpha_{f}(0)$ (fixed)   &   .785000E+00 & .000000E+00 &   .785000E+00 &.000000E+00\\
 $Q^{2}_{f}$ (GeV$^{2}$)   &   .316038E+00 & .566126E-02 &   .329387E+00 &.551927E-02\\
 $d_{f}$                   &   .675356E+00 & .369510E-02 &   .674280E+00 &.361970E-02\\
\hline\hline $\chi^{2}/d.o.f.$ & \multicolumn{2}{c||} {0.892} &
 \multicolumn{2}{c|} {0.925}\\
\hline
\end{tabular}
\end{center}
}
\end{table}
}

In the "full" (\i.e. with BCDMS and SLAC points) regions A and B the model
gives
$$
\mbox{Region A:} \qquad \chi^{2}/d.of.=0.899,
$$
$$
\mbox{Region B:} \qquad \chi^{2}/d.of.=0.948.
$$

We complete, in the kinematical regions where $x\leq 0.1$
$$
\chi^{2}/d.o.f.=0.925 \qquad \mbox{if} \qquad W\geq 6\mbox{ GeV},
$$
$$
\chi^{2}/d.o.f.=0.955 \qquad \mbox{if} \qquad W\geq 3\mbox{ GeV}.
$$

\smallskip
\subsubsection{\bf Comparison between models at small $x$}
Let us briefly discuss the obtained results when $x\leq 0.07$. In order to
make the comparison between models more clear, we collect the
corresponding $\chi^{2}_{d.o.f}$-s in Table~\ref{Tcompar}, where we recall
also some characteristics of the models.

{\renewcommand{\arraystretch}{1.2}%
\begin{table}
\caption{Comparison of the quality of data descriptions at small $x$ in
the 4 investigated models; the kinematical regions are defined in the
text} \label{Tcompar}
{\footnotesize
\begin{center}
\begin{tabular}{|c|c||c|c|}
\hline
 & Pomeron   &\multicolumn{2}{c|}{$\chi^{2}/d.o.f.$}\\
\cline{3-4}
Model of Pomeron & singularity           & Fit  ($W>6$ GeV) &  Fit ($W>3$ GeV) \\
          &                              &    A$_1$; A      &    B$_1$; B      \\
\hline
Soft+Hard Pomeron  & simple poles                     &    1.089  &    1.176    \\
(\ref{DLorF2})-(\ref{DLorFf})  & $\alpha(0)>1$               &    1.097  &    1.504    \\
\hline
Soft Dipole Pomeron       &simple + double poles             &    0.911  &    0.948   \\
(\ref{ampl})-(\ref{B-xp})      & $\alpha(0)=1$               &    0.936  &    1.021  \\
\hline
Modified two-Pomeron     & simple poles                     &   0.959  &    0.996  \\
(\ref{DLmod})-(\ref{DLmodf})&$\alpha(0)\gsim 1,  \alpha(0)=1$&   0.963  &    1.004  \\
\hline
Generalized Logarithmic Pomeron   &simple + double poles     &   0.892  &    0.925  \\
(\ref{LogG})-(\ref{LogG-B})            & $\alpha(0)=1$       &   0.899  &    0.948  \\
\hline
\end{tabular}
\end{center}
}
\end{table}
}

All investigated models well describe the data in the two kinematical
regions. Nevertheless it is clear that the models without a hard Pomeron
(the SDP model and especially the GLP one) are preferable to the original
D-L model, which include a hard Pomeron with $\alpha_{P}(0)>1$.

Thus in our opinion the most interesting and important result which has
been derived from the above comparison of the models is that all SF data
at $x<0.1$ and $Q^{2}\leq 3000$ GeV$^{2}$ are described with a high
quality {\bf without a hard Pomeron}. Moreover, these data support the
idea that the soft Pomeron, either is a double pole with $\alpha_{P}(0)=1$
in the angular momentum $j$-plane or is a simple pole having intercept
$\alpha_{P}(0)=1+\epsilon$ with a very small $\epsilon $. There is no
contradiction with perturbative QCD where BFKL Pomeron has large
$\epsilon$. Firstly, it is well known that the corrections to BFKL Pomeron
are large and the result of their summation is unknown yet. Secondly, the
kinematical region ($x\ll 1,\quad W^{2}\gg Q^{2}$) is a region where the
Regge approach should be valid and where non-perturbative contributions
(rather than perturbative ones) probably dominate.

In fact, we have two soft Pomerons in the SDP and LGP models, the first
one, simple pole located in $j$-plane exactly at $j=1$ and giving a
negative contribution to cross-section. This negative sign is a
phenomenological fact, nevertheless such a term can be treated as a
constant part of the dipole Pomeron rescatterings giving a negative
correction to the single exchange. On the other hand a simple pole with
intercept equal one can be treated as a crossing-even component
three-gluon exchange \cite{3gluon}.

The successful description of small-$x$ domain within the SDP and GLP
models allows us to apply them~\footnote{ We tried also to extend the
Mod2P model to large $x$ by using simple $(1-x)^{B_i(Q^{2})}$ factors. We
failed to get a good agreement with the data.} to the extended region C,
defined by the inequalities (\ref{setC}).

\subsection{Soft Pomeron models at large $x$ }
In this section we present the results of the fits to the extended
$x$-region, up to $x\leq 0.75$, \ie to region C, performed in the Soft
Dipole Pomeron model and in the newly proposed Generalized Logarithmic
Pomeron model. The values of the fitted parameters, their errors as well
as $\chi^{2}$ are given in Table~\ref{T2-all}.

{\renewcommand{\arraystretch}{1.2}%
\begin{table}
\caption{ Parameters obtained from the fit to the data set in region C
((\ref{setC})) within the Soft Dipole Pomeron model (left) and the
Generalized Logarithmic Pomeron model (right).} \label{T2-all}
\medskip
\begin{center}
{\footnotesize
\begin{tabular}{|l|c|c||l|c|c|}
\hline \multicolumn{3}{|c||}{SDP model}&\multicolumn{3}{|c|}{GLP model}\\
\hline
 Parameter  & value & $\pm$error & Parameter  & value & $\pm$error\\
\hline\hline
$C_1$ (GeV$^{-2}$)               &   .210000E-02  &.262020E-05 & $C_{0}$ (GeV$^{-2}$)      &$-$.860438E+00 & .463009E-02\\
$Q_1^2$ (GeV$^2)$                &   .965340E+01  &.126293E-01 & $Q^{2}_{0}$ (GeV$^{2}$)   &   .133405E+01 & .928886E-02\\
$Q_{1d}^2$ (GeV$^2)$             &   .154944E+01  &.490752E-02 & $d_{0}$                   &   .113778E+01 & .308424E-02\\
$d_{1\infty}$                    &   .130005E+01  &.187044E-02 & $Q^{2}_{0b}$ (GeV$^{2}$)  &   .741627E+01 & .249967E+00\\
$d_{10}$                         &   .866015E+01  &.194363E-01 & $b_{0\infty}$             &   .703568E+01 & .602611E-01\\
$Q_{1b}^2$ (GeV$^2)$             &   .315548E+00  &.644633E-02 & $b_{00}$                  &   .142693E+01 & .817529E-01\\
$b_{1\infty}$                    &   .290978E+01  &.676692E-02 &                           &               &            \\
$b_{10}$                         &$-$.205020E+02  &.390882E+00 &                           &               &            \\
\hline\hline
$C_2$ (GeV$^{-2}$)               &$-$.774241E-02  &.819441E-05 & $C_{s}$ (GeV$^{-2}$)      &   .444070E+00 & .200807E-02 \\
$Q_2^2$ (GeV$^2)$                &   .219350E+02  &.234796E-01 & $a $                      &   .143489E+01 & .168555E-01 \\
$Q_{2d}^2$ (GeV$^2)$             &   .300412E+01  &.140016E-01 & $\epsilon $               &   .434764E+00 & .166723E-02 \\
$d_{2\infty}-d_{1\infty}$        &   .000000E+00  &.000000E+00 & $Q^{2}_{s\,0}$ (GeV$^{2}$)&   .188709E+00 & .347041E-02 \\
$d_{20}$                         &   .433861E+01  &.137719E-01 & $d_{s\, 0}$               &   .733135E+00 & .281299E-02 \\
$Q_{2b}^2$ (GeV$^2$)             &   .898304E+01  &.871305E-01 & $Q^{2}_{s\,1}$ (GeV$^{2}$)&   .892069E+00 & .915517E-02 \\
$b_{2\infty}$                    &   .340630E+01  &.455406E-02 & $d_{s\,}$                 &   .693609E+00 & .267361E-02 \\
$b_{20}$                         &   .120264E+01  &.150493E-01 & $Q^{2}_{sb}$ (GeV$^{2}$)  &   .192698E+02 & .135853E+01 \\
                                 &                &            & $b_{s\infty}$             &   .110421E+02 & .258010E+00 \\
                                 &                &            & $b_{s0}$                  &   .312619E+01 & .199118E+00 \\
\hline \hline
$\alpha_f(0)$ (fixed)            &   .785000E+00  &.000000E+00 & $C_{f}$ (GeV$^{-2}$)      &   .211700E+01 & .731039E-02 \\
$C_f$ (GeV$^{-2}$)               &   .277583E-01  &.376481E-04 & $\alpha_{f}(0)$ (fixed)   &   .785000E+00 & .000000E+00 \\
$Q_f^2$ (GeV$^2)$                &   .165653E+02  &.238896E-01 & $Q^{2}_{f}$ (GeV$^{2}$)   &   .901062E+00 & .500511E-02 \\
$Q_{fd}^2$ (GeV$^2)$             &   .384787E+00  &.985244E-03 & $d_{f}$                   &   .863201E+00 & .126160E-02 \\
$d_{f\infty}$                    &   .136494E+01  &.143903E-02 & $Q^{2}_{fb}$ (GeV$^{2}$)  &   .280848E+01 & .891465E-01 \\
$d_{f0}$                         &   .469211E+02  &.111801E+00 & $b_{f\infty}$             &   .354614E+01 & .826982E-02 \\
$Q_{fb}^2$ (GeV$^2)$             &   .819589E+01  &.809316E-01 & $b_{f0}$                  &   .832717E+00 & .610819E-01 \\
$b_{f\infty}$                    &   .332856E+01  &.501372E-02 &                           &               &             \\
$b_{f0}$                         &   .680110E+00  &.146980E-01 &                           &               &             \\
\hline\hline
$\chi^{2}/d.o.f.$ & \multicolumn{2}{c|} {1.053} &$\chi^{2}/d.o.f.$ & \multicolumn{2}{c|} {1.074} \\
\hline
\end{tabular}
}
\end{center}
\end{table}
}

In order to compare the quality of our fits with those obtained in an
other known model, we have performed as an example the same fit in the
ALLM model~\cite{ALLM}. This model incorporates an effective Pomeron
intercept depending on $Q^{2}$ and cannot be considered as a Regge-type
model. Nevertheless, it leads to a quite good description of the data in
the same kinematical region: we obtained $\chi^{2}/d.o.f.\approx 1.11$ by
limiting the intercept of $f$-Reggeon to a reasonable lower bound
$\alpha_{f}(0)=0.5$.

\begin{figure}[ht]
\begin{center}
\includegraphics[scale=0.55]{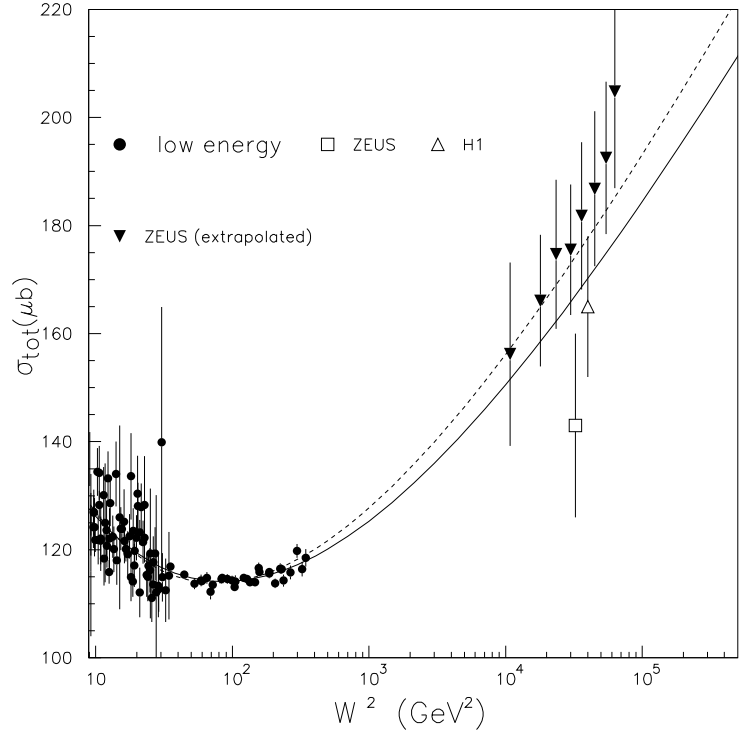}
\caption{Total $\gamma p$ cross-section versus $W^2$ in SDP model (solid
line) and in GLP model (dashed line). Data of \cite{br99} extracted from
the SF at low $Q^{2}$ by the Zeus collaboration are also shown in the
figure but not included in the fit.} \label{FSig}
\end{center}
\end{figure}

\begin{figure}
\begin{center}
\includegraphics*[scale=0.9]{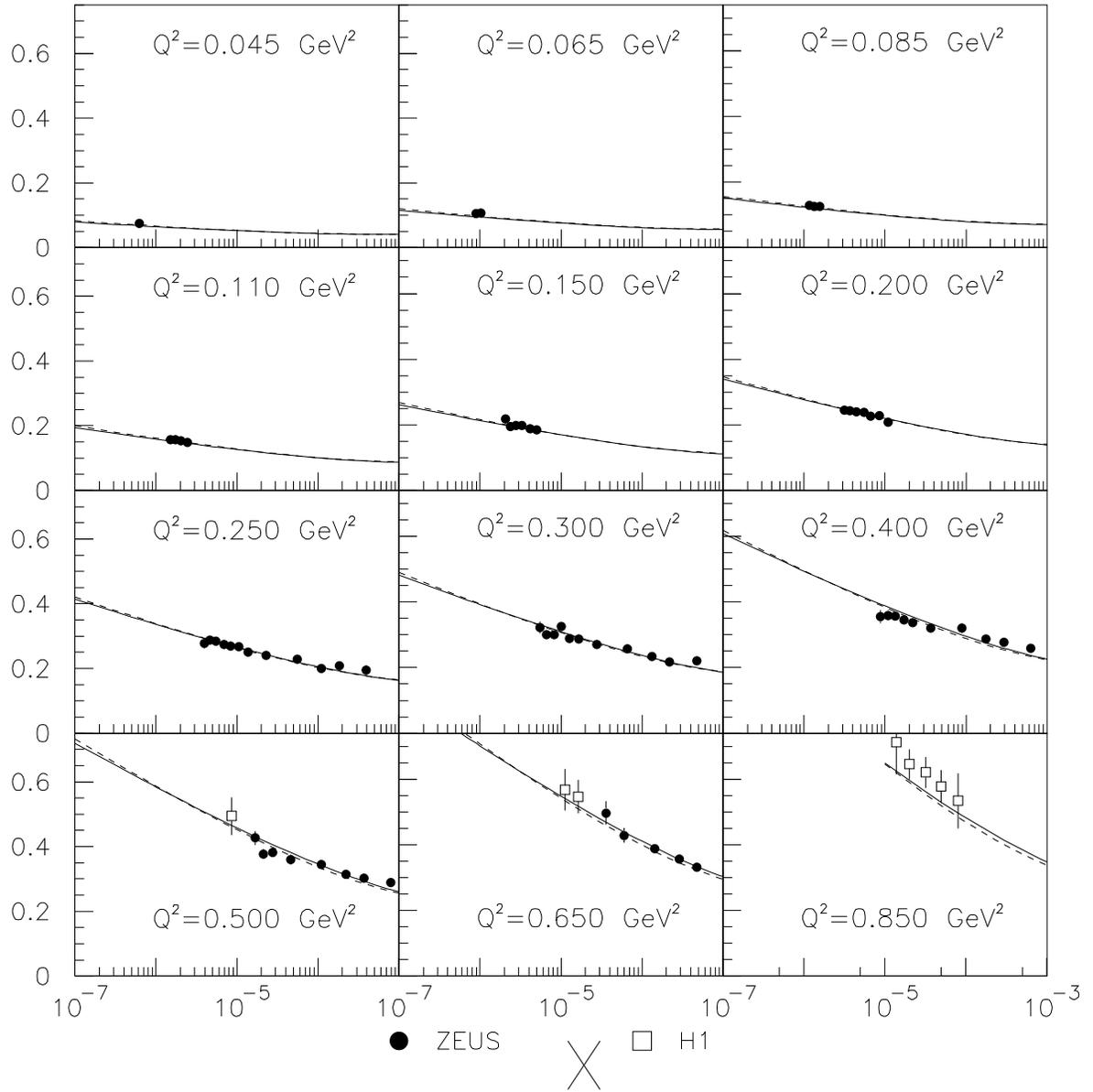}
\caption{Structure function at small $Q^2$ versus $x$. Solid line is $F_2$
calculated
within SDP model, dashed line is $F_2$ within GLP model.} \label{FF2smallQ}
\end{center}
\end{figure}

\begin{figure}
\begin{center}
\includegraphics[scale=0.9]{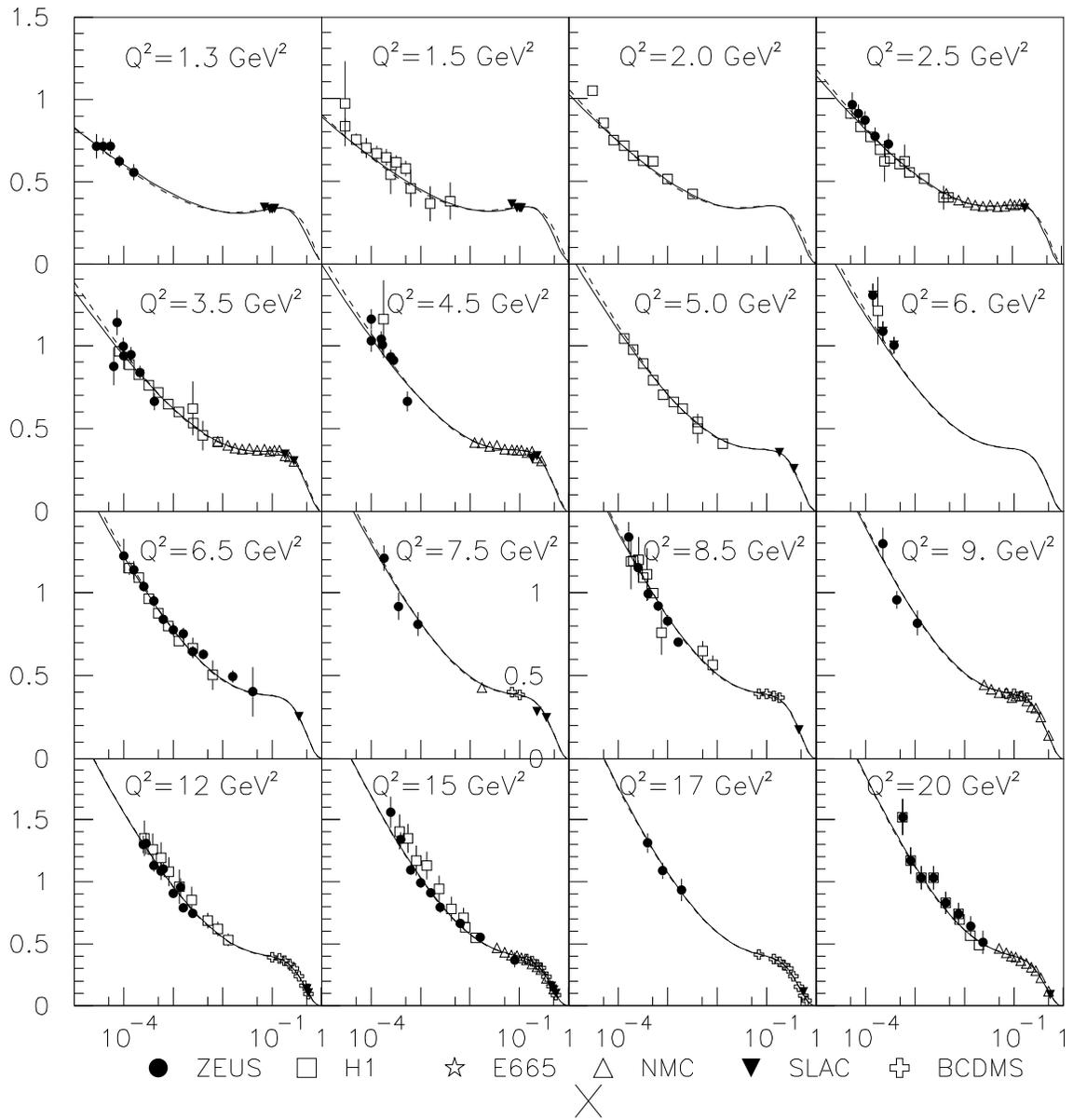}
\caption{Same as in Fig.~\ref{FF2smallQ} for intermediate $Q^{2}$}
\label{FF2interQ}
\end{center}
\end{figure}

\begin{figure}
\begin{center}
\includegraphics[scale=0.9]{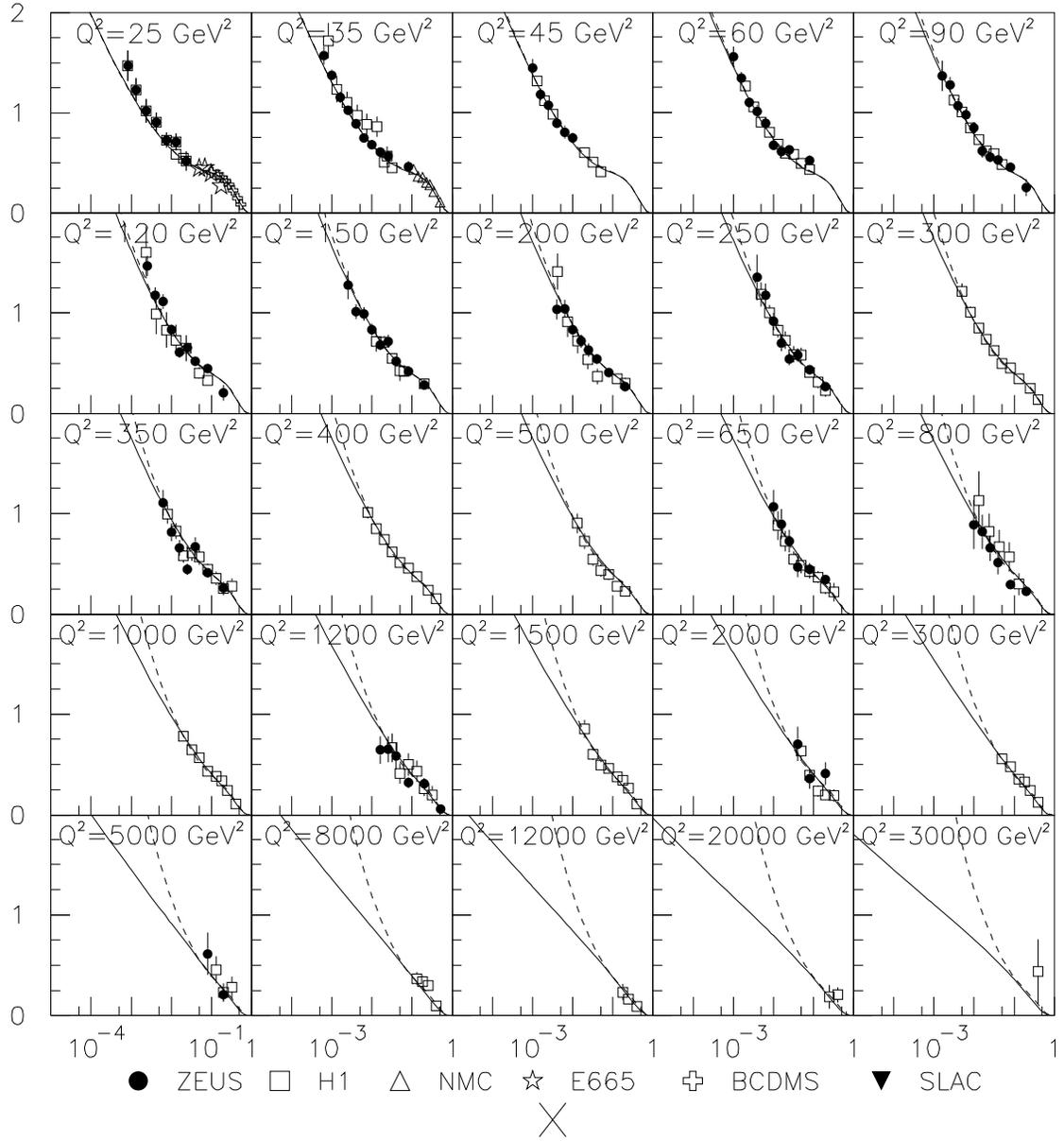}
\caption{Same as in Fig.~\ref{FF2smallQ} for large $Q^{2}$. The data
represented in the lower row of icons, at $Q^{2}\geq $ 5000 GeV$^2$, are not
included in the fit}
\label{FF2largeQ}
\end{center}
\end{figure}

 The behaviour of the theoretical curves for the cross-section
$\sigma_{tot}^{\gamma p}$ versus the center of mass energy squared and for
the proton structure function $F_{2}$ versus $x$ for $Q^2$ ranging from
the lowest to the highest values is shown in
Figs.~\ref{FSig}-\ref{FF2largeQ} for both models.

\bigskip
One can see from the figures that
\begin {itemize}
\item
both calculated $\gamma p$ cross-sections are above the two experimental
HERA results at high energy; rather, they would be in agreement with the
extrapolation performed~\cite{br99} from very low $Q^2$. The GLP model
reveals a steeper rise with the energy than the SDP model.
\item
The calculated SDP and GLD proton structure functions can be distinguished
by eye only outside the fitted range, especially at high $Q^2$ where the
steeper rise of GLP model is evident.
\item
The SF curves calculated in the GLP model have a larger curvature
(especially at high $Q^{2}$) than we expected and consequently larger
logarithmic derivatives $B_{x}=\partial \ell nF_{2}(x,Q^{2})/\partial \ell
n(1/x)$~\footnote{ A comparative detailed investigation of the derivatives
of the proton structure with respect to $x$ and $Q^{2}$ is under
progress.}.
\end{itemize}
The last feature is reflected in the partial $\chi^{2}$ for different
intervals of $Q^{2}$, as it can be seen in Table~\ref{TpartQ}, where we
compare the quality of the data description in such intervals. Indeed the
GLP model "works" better in the region of intermediate $Q^{2}$, while SDP
model describes better the data at small and large $Q^{2}$ (including data
on the total real $\gamma p$ cross-section). A similar analysis made for
intervals in $x$ would show that SDP model is more successful in region of
small and large values of $x$ and GLP model is for intermediate $x$, in
agreement with the fact that the available data at intermediate $Q^{2}$
have also intermediate values of $x$.

\begin{table}[ht]
\caption{Partial values of $\chi^{2}$ for different intervals of
$Q^{2}$ in SDP and GLP models.}\label{TpartQ}

\medskip
\begin{center}
\begin{tabular}{|l|c|c|c|}
\hline
 Interval of $Q^2$ (GeV$^2$) & Number of points & SDP model & GLP model \\
\hline\hline
         $Q^2$=0             &        99        &  121.96   &  133.15   \\
\hline
        0$<Q^2\leq$5         &       404        &  325.30   &  348.32   \\
\hline
        5$<Q^2\leq$50        &       540        &  621.81   &  614.73   \\
\hline
       50$<Q^2\leq$100       &       101        &  112.47   &   94.27   \\
\hline
      100$<Q^2\leq$500       &       150        &  141.79   &  151.62   \\
\hline
      500$<Q^2\leq$3000      &        96        &  116.38   &  126.87   \\
\hline
\end{tabular}
\end{center}
\end{table}
\section{Conclusion}
First of all, we would like to emphasize once more two important points.

1). The kinematical regions A (or A$_{1}$) and B (or B$_{1}$) where $x$ is
small are the domains where all conditions to apply the Regge formalism
are satisfied~: $W^{2}\gg m_{p}^{2},\quad W^{2}\gg Q^{2},\quad x\ll 1$.
However because of universality of Reggeons and of existing correlations
between Pomeron and $f$-Reggeon contributions, it is important to fix
$\alpha_{f}(0)$ to the value determined from the hadronic data on
resonances and on elastic scattering.

2). Analyzing the ability of any model to describe the data, it is
necessary to verify how important are the assumptions on which the model
is based. A possible mean holds in comparing the original model with an
alternative one constructed without such assumptions (of course using a
common set of experimental data).

In this work, we respect these two points and our conclusions are
the following.

{\bf Small} $\mathbf x$. We have shown that the available data can
be described without a hard Pomeron component. Moreover the models without a
hard Pomeron lead to a better description of data (by $\approx $10\% in
terms of $\chi^{2}$). Furthermore, the best description is obtained in a
model where the two Pomeron components have the trajectories with
an intercept one.

We have proposed a new model for the proton structure function: the
"Generalized Logarithmic Pomeron" model, which has not a hard Pomeron, but
mimics its contribution at large $Q^{2}$. In the region of small $x$ this
model gives the best $\chi^{2}/d.o.f.$

{\bf Small and large} $\mathbf x$. Multiplying each $i$-component of the
Soft Dipole Pomeron and of the Generalized Logarithmic Pomeron models by a
factor $(1-x)^{B_{i}(Q^{2})}$, we can describe well not only small-$x$
data but also data at all $x\leq 0.75$. As noted recently~\cite{D-Lnew},
these factors can be considered as an effective contribution of all
daughter trajectories associated with Pomeron and $f$-Reggeon. Thus, their
introduction is only an extension of the Regge approach to the whole
kinematical $x$-region.

In spite of almost equivalent qualities of description, a precise analysis
shows that these two models differently describe the data in the different
regions of $x$ and $Q^{2}$. The extended SDP model is more successful at
small $x$, while the extended GLP model better describes the data at
intermediate $Q^{2}$ and $x$. It would be interesting to construct a model
incorporating the best features of both.

Concluding, we stress again that the available data on the proton structure
function and on the $\gamma p$ cross-section do not yield
explicit indications in favor of an existing hard Pomeron.

\bigskip
{\it Acknowledgements} We have the pleasure to thank M. Giffon for a
critical reading of the manuscript.

\bigskip


\begin{thebibliography}{99}
\bibitem{Collins}
P.D.B. Collins, {\it Introduction to Regge Theory and High Energy
Physics}, Cambridge University Press (1977).

\bibitem{GLR-83}
L.V. Gribov, E.M. Levin, M.G. Ryskin, Phys. Rep. {\bf 100}, 1 (1983).

\bibitem{ALLM}
H. Abramovicz \etal, Phys. Lett. B {\bf 269} 465, (1991);

H. Abramovicz, A. Levy, DESY 97-251, hep-ph/9712415 (1997).

\bibitem{BK}
B. Badelek, J. Kwiecinski, Phys. Lett. B {\bf 295}, 263 (1992); Rev. Mod.
Phys. {\bf 68}, 445 (1996).

\bibitem{Be-95}
M. Bertini \etal, in "Strong interactions at long distances", edited by L.
Jenkovszky, Hadronic Press, Palm Harbor, FL U.S.A, p.181, (1995).

\bibitem{bpr}
A. Bialas, R. Peschanski, Ch. Royon, Phys. Rev. D {\bf 57} 6899 (1998).

\bibitem{D-L2pom}
A. Donnachie, P.V. Landshoff, Phys. Lett. B {\bf 437}, 408 (1998).

\bibitem{DLM-dpom}
P. Desgrolard, S. Lengyel, E. Martynov, Eur. Phys. Jour. C {\bf 7}, 655
(1999); LYCEN 98102, hep-ph/9811380 (1998).

\bibitem{djp}
P. Desgrolard, L. Jenkovszky, F. Paccanoni, Eur. Phys. Jour. C {\bf 7},
263 (1999).

\bibitem{djlp}
P. Desgrolard, L. Jenkovszky, A. Lengyel, F. Paccanoni, Phys. Lett. B {\bf
459}, 265 (1999).

\bibitem{PetPro}
V.A. Petrov, A.V. Prokudin, hep-ph/9912248 (1999).

\bibitem{NSZ}
N.N. Nikolaev, J. Speth, V.R. Zoller, Phys. Lett.
B {\bf 473}, 157 (2000).

\bibitem{Kaidalov}
A.B. Kaidalov, hep-ph/0103011 (2001) and
references therein.

\bibitem{Schild}
D. Schildknecht, Nucl. Phys. (Proc.Suppl.) B {\bf 99}, 121 (2001);

G. Cvetic, D. Schildknecht, B. Surrow, M. Tentyukov, hep-ph/0102229
(2001).

\bibitem{TroTyu}
S.M. Troshin, N.E. Tyurin, hep-ph/0102322 (2001).

\bibitem{D-Lnew}
A. Donnachie, P.V. Landshoff, hep-ph/0105088 (2001).

\bibitem{ChWu}
H. Cheng, T.T. Wu, Phys. Rev. D {\bf 1}, 2775 (1970);
Phys. Rev. Lett. {\bf 24}, 1456 (1970).

\bibitem{BFKL}
 L.N. Lipatov, Sov. J. Nucl. Phys. {\bf 23}, 338 (1976);
E.A. Kuraev, L.N. Lipatov, V.S. Fadin, Sov. Phys. JETP {\bf 45}, 199
(1977); Y.Y. Balitsky and L.N. Lipatov, Sov. J. Nucl. Phys. {\bf 28}, 822
(1978).

\bibitem{FadLip}
V.S. Fadin, L.N. Lipatov, Phys. Lett. B {\bf 429}, 127 (1998) and
references therein.

\bibitem{D-Lhadr}
A. Donnachie, P.V. Landshoff, Phys. Lett. B {\bf 296}, 227 (1992).

\bibitem{DGLM}
P. Desgrolard, M. Giffon, A. Lengyel, E. Martynov,
Nuovo Cim. A {\bf 107}, 637 (1994).

\bibitem{CEKLT}
J.R. Cudell et al., Phys. Rev. D {\bf 61}, 034019 (2000).

\bibitem{DGMP-ed}
P. Desgrolard, M. Giffon, E. Martynov, E. Predazzi,
Eur. Phys. J. C {\bf 18}, 555 (2001).

\bibitem{ad00}
H1 collaboration, C. Adloff \etal , DESY-00-181, hep-ex/0012053 (2000).

\bibitem{br00}
ZEUS collaboration, J. Breitweg \etal , Nucl. Phys. B {\bf 487}, 53
(2000).

\bibitem{ah95}
H1 collaboration, T. Ahmed \etal , Nucl. Phys. B {\bf 439}, 471 (1995).

\bibitem{ai96}
H1 collaboration, S. Aid \etal , Nucl. Phys. B {\bf 470}, 3 (1996).

\bibitem{ad97}
H1 collaboration, C. Adloff \etal , Nucl. Phys. B {\bf 497}, 3 (1997).

\bibitem{ad99}
H1 collaboration, C. Adloff \etal , Eur. Phys. J. C {\bf 13}, 609 (2000).

\bibitem{de96}
ZEUS collaboration, M. Derrick \etal , Zeit. Phys. C {\bf 72}, 399 (1996).

\bibitem{br97}
ZEUS collaboration, J. Breitweg \etal , Phys. Lett. B {\bf 407}, 432
(1997).

\bibitem{br99}
ZEUS collaboration, J. Breitweg \etal , Eur. Phys. J. C {\bf 7}, 609
(1999).

\bibitem{ar97}
NMC collaboration, M. Arneodo \etal , Nucl. Phys. B {\bf 483}, 3 (1997).

\bibitem{ad96}
E665 collaboration, M.R. Adams \etal , Phys. Rev. D {\bf 54}, 3006 (1996).

\bibitem{slac}
L. W. Whitlow (Ph. D. thesis), SLAC-PUB 357 (1990);

SLAC old experiments, L. W. Whitlow \etal , Phys. Lett. B {\bf 282}, 475
(1992).

\bibitem{be89}
BCDMS collaboration, A. C. Benvenuti \etal , Phys. Lett. B {\bf 223}, 485
(1989).

\bibitem{pdg}
The data on the total $\gamma p$ cross-section are extracted from
http://pdg.lbl.gov.

ZEUS collaboration, M. Derrick \etal, Zeit. Phys. C {\bf 63}, 391 (1994);

H1 collaboration, S. Aid \etal , Zeit. Phys. C {\bf 69}, 27 (1995).

\bibitem{MAQM} P. Desgrolard, M. Giffon, E. Martynov, E. Predazzi,
Eur. Phys. J. C {\bf 9}, 623 (1999).

\bibitem{DLM-bx}
P. Desgrolard, A. Lengyel, E. Martynov, Nucl. Phys. (Proc. Supp.) B {\bf
99}, 168 (2001).

\bibitem{3gluon}
L.N. Lipatov, Nucl. Phys. (Proc. Suppl.) B {\bf 79}, 207 (1999);

J. Bartels, L.N. Lipatov, G.P. Vacca, Phys. Lett. B {\bf 477}, 178 (2000).

\end{thebibliography}
\end{document}